\documentclass[aps,pra,preprint,groupedaddress,showpacs]{revtex4-1}
\usepackage{amssymb}
\usepackage{amsmath}
\usepackage{dcolumn}
\usepackage{bm}
\usepackage{graphicx}
\usepackage{mathrsfs}

\usepackage{color}
\usepackage{ulem}

\newcommand{\pei}[1] {\textcolor{black}{#1}}

\input{tcilatex}

\begin{document}

\title{Super sub-wavelength patterns in photon coincidence detection}
\author{Ruifeng Liu}
\author{Pei Zhang}
\email{zhangpei@mail.ustc.edu.cn}
\author{Yu Zhou}
\author{Hong Gao}
\author{Fuli Li}
\email{flli@mail.xjtu.edu.cn}

\begin{abstract}
High-precision measurements implemented by means of light is desired in all fields of science. However, light is a wave and Rayleigh
criterion gives us a diffraction limitation in classical optics which restricts to get arbitrary \pei{high} resolution. Sub-wavelength interference
has a potential application in lithography to beat the classical Rayleigh limit of resolution. We carefully study the second-order correlation theory
to get the physics behind sub-wavelength interference in photon coincidence detection. A Young's double-slit experiment with pseudo-thermal light is carried
out to test the second-order correlation pattern. The result shows that when different scanning ways of two point detectors are chosen,
\pei{one} can get super sub-wavelength interference patterns. We then give a theoretical explanation to this surprising result, and find this
explanation is also suitable for the result by using entangled light. Furthermore, we discuss the limitation of this kind of super sub-wavelength
interference patterns in quantum lithography.

\bigskip
\bigskip

\end{abstract}

%\pacs{42.50.Dv, 03.67.-a, 76.30.Mi}
\maketitle

\address{MOE Key Laboratory for Nonequilibrium Synthesis and
Modulation of Condensed Matter, and Department of Applied Physics, Xi'an
Jiaotong University, Xi'an 710049, People's Republic of China}

\noindent \textbf{Introduction}

Quantum lithography was first proposed by Boto \textit{et al} \cite{Boto2000} that allows \pei{an} N-times smaller spacing of interference fringes than the
classical Rayleigh limit  \cite{Rayleigh} of spatial resolution. To date, a true laboratory verification of quantum lithography is still extremely challenging due to
the absence of N-photon absorption lithographic materials \cite{Boyd2012}. The alternative way to demonstrate quantum lithography in experiment
is using N detectors operating in coincidence to mimic the effect of a true N-photon absorbing resist.
In interferometric lithography, coherent light beams shine on a mask and form an interference pattern. If the spacing of the interference fringes is smaller than the size of wavelength $\lambda$ \cite{explain}, which is  the classical Rayleigh limit \cite{Rayleigh}, one obtains so called a sub-wavelength interference pattern, which is expected to be used in lithography.
The sub-wavelength interference pattern has not only been obtained using entangled light~\cite{D'Angelo2001} by simultaneously scanning
the two point detectors in same directions but also observed using thermal light~\cite{Scarcelli2004,Xiong2005,Zhai2005} by simultaneously
scanning in opposite directions. However, the question of why the sub-wavelength interference pattern can be \pei{obtained only when the two point detectors are scanned in those ways}
%we scan two point detectors like \pei{those} 
is not very clear, and is there any other better scanning way to get higher resolution than $\lambda/2$? In our study, we find that the scanning way is not unique. 
\pei{By choosing propers canning ways, even super sub-wavelength interference pattern can be achieved.}
%By choosing different scanning ways, we can get super sub-wavelength interference pattern. 
This surprising result makes us to think about the physics behind this phenomenon and whether we can achieve an interference
pattern of arbitrary high resolution. We will address these questions in this manuscript \pei{and give a suggestion on how to get the full picture of second-order intensity correlation}.

In light interference and diffraction experiments, the observed quantity is the first-order correlation function of light on a detecting plane at the far-field region,
\begin{equation}
G^{(1)}(\textbf{r},t)=Tr[\rho E^{(-)}(\textbf{r},t) E^{(+)}(\textbf{r}, t)],
\label{eq:1}
\end{equation}
where $E^{(-)}(\textbf{r}, t)$ and $E^{(+)}(\textbf{r}, t)$ are operators for the \pei{negative and positive frequency} parts of the light field,
$\rho$ is the density matrix operator of the light source, and $\textbf{r}$ is a transverse coordinate vector on the detecting plane.
The measured first-order correlation Eq.~(\ref{eq:1}) directly represents the intensity distribution of light on the detecting plane.
%\\
%\\

%\noindent \textbf{Second-order correlation.} 
Besides the first-order correlation, light can have the second-order correlation which is defined as
\begin{eqnarray}
&&G^{(2)}(\textbf{r}_{1}, t_{1}; \textbf{r}_{2}, t_{2})=Tr[\rho E^{(-)}(\textbf{r}_{1}, t_{1})E^{(-)}(\textbf{r}_{2}, t_{2})  \notag\\
&&\ \ \ \ \ \ \ \ \ \ \ \ \ \ \ \ \ \ \ \ \ \ \ \ \ \ E^{(+)}(\textbf{r}_{2}, t_{2})E^{(+)}(\textbf{r}_{1}, t_{1})].
\label{eq:2}
\end{eqnarray}
In 1956, Hanbury-Brown and Twiss (HBT) \cite{HBT19561,HBT19562} first observed the second-order intensity correlation phenomenon of light in astronomy. This \pei{investigation arouse research interest to the second-order correlation property of light \cite{HBT1974,mandel}, and inspired Glauber's work on quantum optics \cite{Glauber1963,Glauber1963pr}.} The HBT experiment can be physically explained by either the classical statistical correlation of the intensity fluctuations \cite{HBT1974} or two-photon (multi-photon) probability amplitude interference \cite{Scarcelli2006}. Although the debate how to understand the physical mechanism behind the HBT experiment exists for a long time, the second-order correlation property of light has been widely applied in various fields such as nonlocal imaging and interference \cite{Pittman1995,Cheng2004,Gatti2004,QI2012,phs94,ejs99,EJS99}, sub-wavelength interference \cite{D'Angelo2001,Scarcelli2004,Xiong2005,Zhai2005} and quantum lithography \cite{Sciarrino08,Tsang2009,Guerrieri2010,Boyd2012}.
\\
\\
\pei{\noindent \textbf{Results}}

\begin{figure}[hbt]
\centering
\includegraphics[width=75mm]{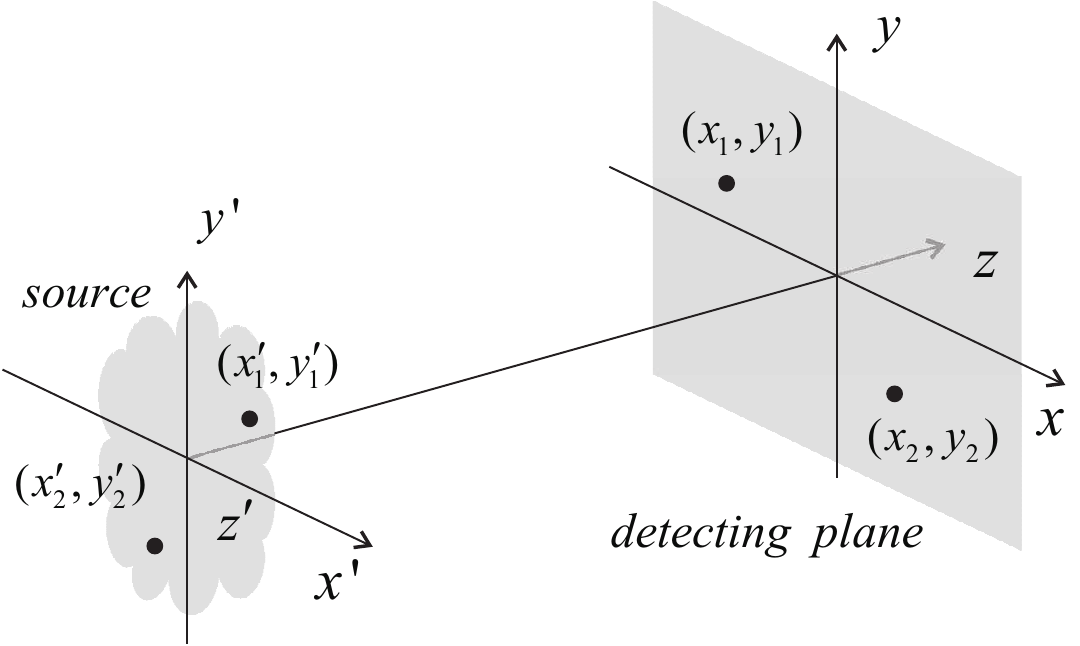}
\caption{(Color online) Schematic of the \pei{correspondence between} source and detecting plane.} 
\label{fig00}
\end{figure}

\noindent In Fig. \ref{fig00}, we show the point-to-point corresponding relation between the source and the detecting plane, which can be described by the impulse response function
$h(\textbf{r},\textbf{r}')$.
If the \pei{source is thermal light}, the second-order correlation function on a detecting plane can be written as \cite{Cheng2004}
\begin{eqnarray}
&&G^{(2)}(\textbf{r}_{1},\textbf{r}_{2})\propto|\int \int d\textbf{r}^{\prime}_{1}
d\textbf{r}^{\prime}_{2}h^{*}(\textbf{r}_{1},\textbf{r}'_{1})h(\textbf{r}_{2},\textbf{r}'_{2})    \notag\\
&&\ \ \ \ \ \ \ \ \ \ \ \ \ \ \ \ \ \ \ \ \ \ \ \ \ \ \ast\langle E^{(-)}(\textbf{r}'_{1})E^{(+)}(\textbf{r}'_{2})\rangle|^{2}.
\label{eq:3}
\end{eqnarray}
Under the condition of paraxial approximation and the assumption that the source is covered by an object with a transmission function $T(\textbf{r}')$, the impulse response function can be expressed as
\begin{eqnarray}
&&h(\textbf{r},\textbf{r}')
=\frac{e^{ikz}}{i\lambda
z}e^{-i\frac{k}{z}\textbf{r}\cdot\textbf{r}'}T(\textbf{r}'),
\label{eq:4}
\end{eqnarray}%
where $k=2\pi/\lambda$ is the wave number. If the light source is \pei{fully incoherent, such as thermal light}, the first-order correlation function of light just emitted from the source $\langle E^{(-)}(\textbf{r}'_{1})E^{(+)}(\textbf{r}'_{2})\rangle$ can have the form $n\delta(\textbf{r}'_{1}-\textbf{r}'_{2})$, where $\delta(\textbf{r})$ is the Dirac delta function, and  $n$ is the mean photon number. Here, we have assumed that the intensity distribution of the incoherent
 light on the source plane is uniform \cite{Scarcelli}. Then the second-order correlation function Eq.~(\ref{eq:3}) can be written in the form
\begin{eqnarray}
&&G^{(2)}(\textbf{r}_{1},\textbf{r}_{2})\propto|\widetilde{T^{2}}(\frac{k}{z}(\textbf{r}_{2}-\textbf{r}_{1}))|^{2},
\label{eq:5}
\end{eqnarray}%
where \pei{$\widetilde{T^{2}}(\textbf{r})$} is the Fourier transformation of
$T^{2}(\textbf{r})$. This result shows that a well-defined Fourier-transform image of the transmittance of \pei{an} object
can be extracted from the second-order correlation function.

The expression Eq.~(\ref{eq:5}) clearly shows that the second-order correlation function is a four-variable function of transverse coordinates $(x_{1},y_{1})$ and $(x_{2},y_{2})$ of the detecting plane. Simple but without loss of generality, one-dimensional object such as double-slit is usually employed to make the correlation function become a two-variable function.
For a one-dimensional symmetric double-slit placed along the y-axis, the correlation function is irrelative to the coordinates $y_{1}$ and $y_{2}$. In this way, the second-order correlation function Eq.~(\ref{eq:5}) can be further simplified to the form
\begin{eqnarray}
&&G^{(2)}(x_{1},x_{2})\propto|\widetilde{T^{2}}(\frac{k}{z}(x_{2}-x_{1}))|^{2}.
\label{eq:6}
\end{eqnarray}%
%Now it becomes two-dimensional.

\begin{figure}[hbt]
\centering
\includegraphics[width=75mm]{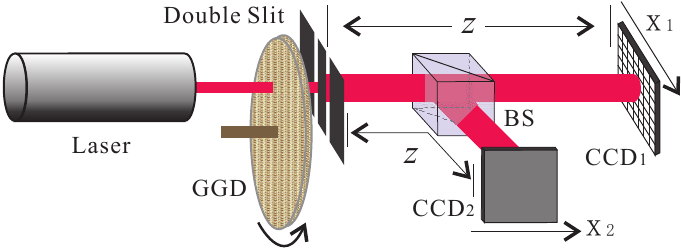}
\caption{(Color online) \pei{Schematic of second-order correlation measurement of double-slit with  charge coupled devices (CCD).} The GGD and double-slit are put as close as possible. The two CCDs are placed at the Fraunhofer diffraction region with respect to  the double-slit. The CCD is $1040\times1392$ array of $4.65\times4.65$ $\rm{\mu m^{2}}$ pixels, and the measurement is made with an exposure time of $0.2$ ms.} \label{setup}
\end{figure}

\noindent \textbf{Experimental test with  pseudo-thermal light.} The experimental setup is shown in Fig.~\ref{setup}. The \pei{pseudo-thermal} light is produced by a semiconductor laser of wavelength $\lambda=457$ nm and a
rotating ground glass disk (GGD). The angular velocity of the GGD is kept at $\omega=\pi$ rad/s. A double-slit is placed right
after the GGD. The width of each slit is $a=0.038$ mm, and the distance between centers of the two slits is $d=0.12$ mm. The diffracted light after the double-slit is split into two parts at a beam splitter (BS) and then recorded by two charge coupled devices (CCD). Both of the CCDs are located at $z=23$ cm behind the double-slit, which means that the CCDs  are placed in the Fraunhofer diffraction region \pei{with respect} to the double-slit.

\begin{figure}[hbt]
\centering
\includegraphics[width=60mm]{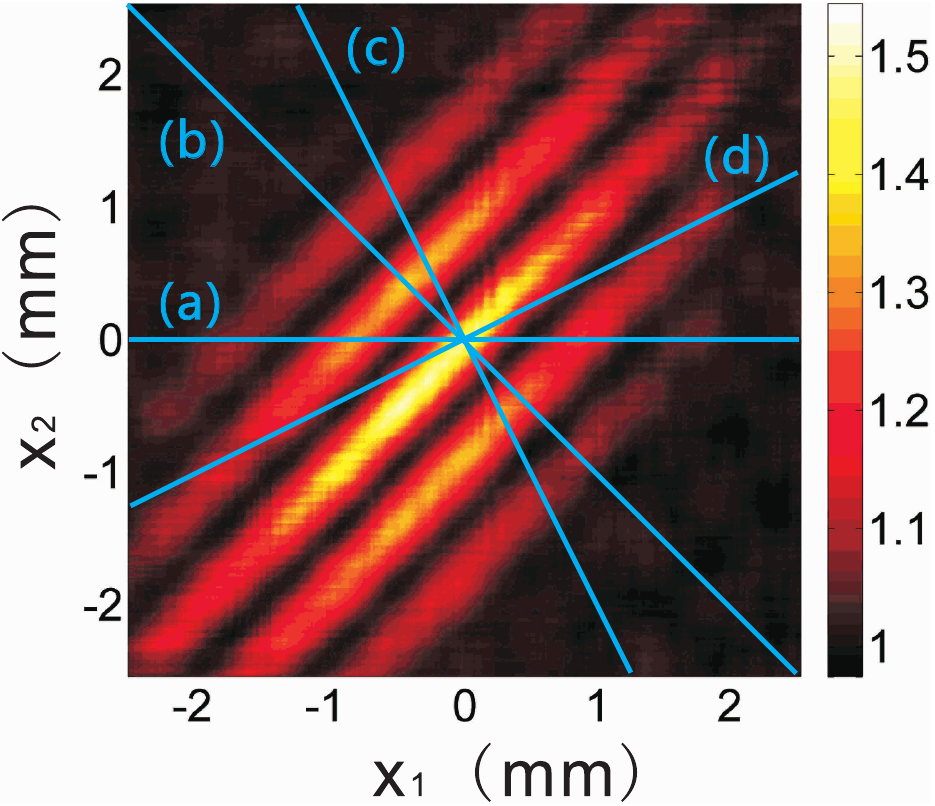}
\caption{(Color online) Second-order correlation function Eq.~(\ref{eq:6}) recorded on the two CCDs. $x_{1}$ and $x_{2}$ are the horizontal ordinates on CCD$_{1}$ and CCD$_{2}$, respectively. The lines represent various scanning ways of the detectors: line (a) for $x_{1}=x$ and $x_{2}=0$; line (b) for $x_{1}=x$ and $x_{2}=-x$; line (c) for $x_{1}=x$ and $x_{2}=-2x$; line (d) for $x_{1}=x$ and $x_{2}=x/2$. } \label{g2}
\end{figure}

As shown above, the second-order correlation function Eq.~(\ref{eq:6}) describes the correlation between arbitrary two points on CCD$_{1}$ and CCD$_{2}$. Here we extract two relevant row data from the two CCDs, and build the second-order correlation function. The result is shown in Fig.~\ref{g2}, where $x_{1}$ and $x_{2}$ are the horizontal coordinates on CCD$_{1}$ and CCD$_{2}$, respectively. Here we use color to represent the \pei{value} of the  second-order correlation function. \pei{It is the two-dimensional second-order interference pattern of the double-slit.}
%This figure shows a two-dimensional interference pattern of the double-slit.

\begin{figure}[hbt]
\centering
\includegraphics[width=75mm]{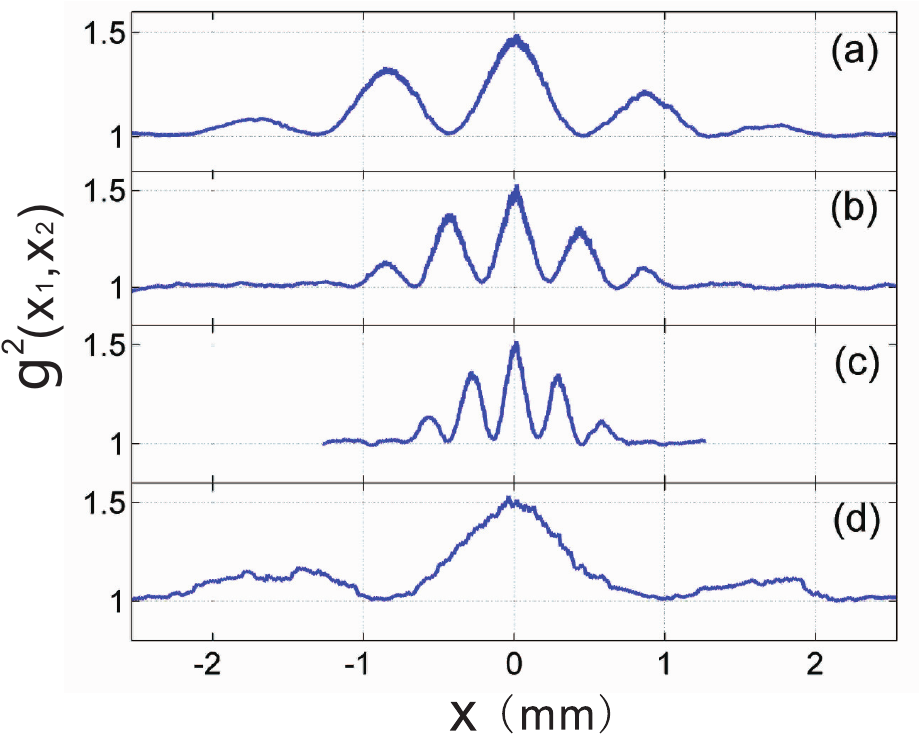}
\caption{(Color online) Cross-sectional curves of (a), (b), (c) and (d) represent the corresponding \pei{lines} in Fig.~\ref{g2}.} \label{g2scan}
\end{figure}

One can obtain a cross-sectional curve with arbitrary spatial resolution by choosing a proper line on the two-dimensional figure. Line (a) in Fig.~\ref{g2} describes the case when one point detector on CCD$_{2}$ plane is fixed and another point detector on CCD$_{1}$ plane is moved along the $x_{1}$ direction. \pei{The curve obtained along this line, as shown in Fig.~\ref{g2scan}(a), is completely same as a traditional Young's interference pattern.} The distance from the zeroth-order peak to the first-order peak in Fig.~\ref{g2scan}(a) is $0.89$ mm which is in coincidence with the theoretical value $0.88$ mm. Scanning with two point detectors in the opposite directions synchronously, which corresponds to line (b) in Fig.~\ref{g2}, we can get \pei{an interference} pattern with $0.44$ mm peak spaces as shown in Fig.~\ref{g2scan}(b). This result is so called the sub-wavelength interference pattern in Ref.~\cite{Scarcelli2004,Xiong2005}.

\begin{figure}[hbt]
\centering
\includegraphics[width=80mm]{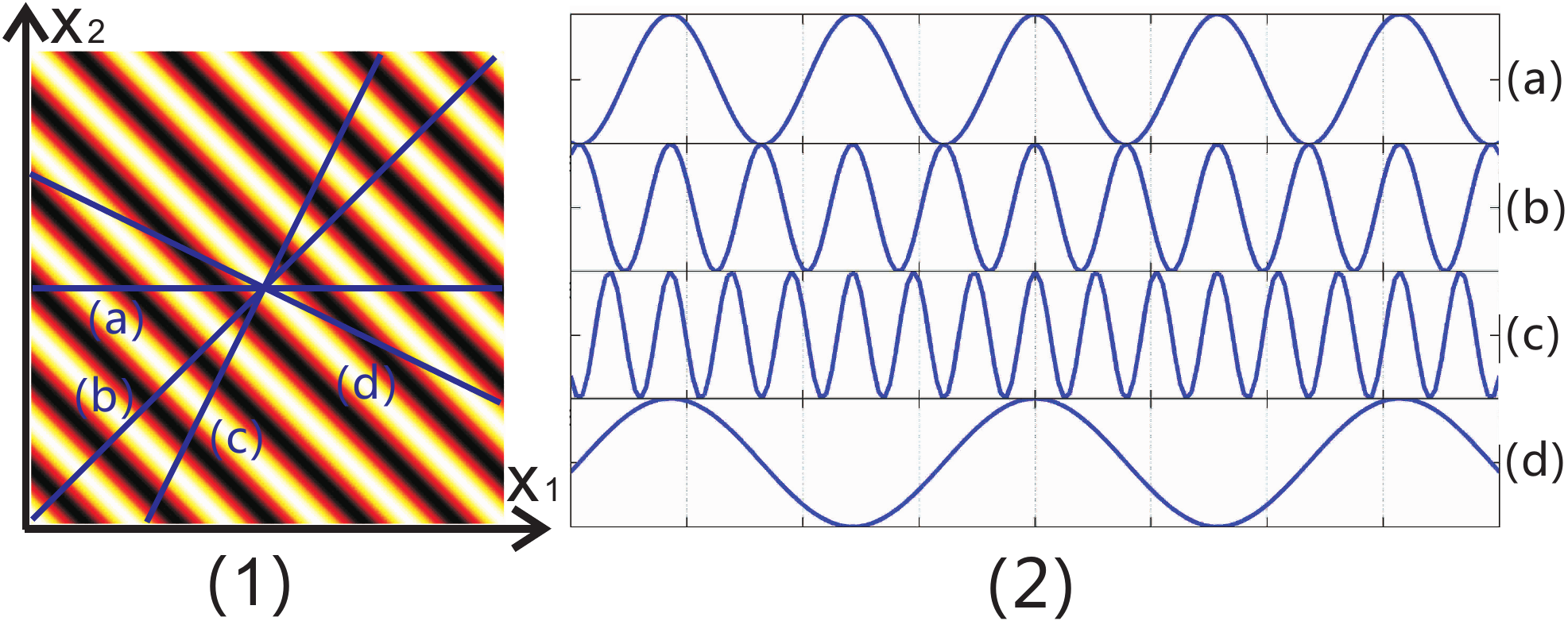}
\caption{(Color online) (1) The full second-order correlation function Eq.~(\ref{eq:7}). (2) (a), (b), (c) and (d) represent the cross-sectional curves which are corresponding to the labelled lines in (1).} \label{entanglement}
\end{figure}

It is obvious that the scanning method is not limited by only the above two conditions. We can choose any scanning method which corresponds to a line in Fig.~\ref{g2} to get a diffracted pattern of arbitrary-wavelength resolution. For example, along line (c) shown in Fig.~\ref{g2} representing a scanning method with $x_{1}=x$, and $x_{2}=-2x$, we obtain a narrower interference pattern than the sub-wavelength one as shown in Fig.~\ref{g2scan}(c). The peak spacing in Fig.~\ref{g2scan}(c) is $0.29$ mm which corresponds to a $\lambda/3$ spatial resolution. If we set $x_{1}=x$, and $x_{2}=x/2$ represented by line (d) in Fig.~\ref{g2}, the resulting data is plotted in Fig.~\ref{g2scan}(d), displaying the $2\lambda$  spatial resolution. The experimental results obtained here well fit with the theoretical prediction Eq.~(\ref{eq:6}). From the above discussion, we come to the conclusion that one-dimensional cross-sectional curves of the second-order correlation function can have the super-sub-wavelength behavior if the scanning way of the detectors is properly chosen.
\\
\\
\noindent \textbf{Results for entangled light source.} \pei{The above conclusion can also be applied to} the case of sub-wavelength interference with entangled light source. In this case, Angelo \textit{et al} \cite{D'Angelo2001} showed that a spatial resolution enhanced interference pattern can be built, which was predicted by Boto \textit{et al} \cite{Boto2000}. In their experiment, a double-slit is placed right after the nonlinear crystal to make sure that the two-photon is generated simultaneously on the up-slit or the down-slit. \pei{They use two point detectors moving in the same direction synchronously and operating in coincidence to mimic} the effect of a true two-photon absorbing resist. They found that the two-photon double-slit interference pattern is 2 times narrower than the traditional pattern with a coherent light. If one assumes that the entangled photon pairs generated simultaneously at either the up or down slit propagate to the detecting plane independently, \pei{the distribution of coincidence counts} at the detecting plane can be approximately  expressed in the form \cite{Steuernagel2004,Tsang2007,Kothe2011}
\begin{eqnarray}
&&G^{(2)}(x_{1},x_{2})\propto \cos^{2}[\frac{kd}{2z}(x_{1}+x_{2})].
\label{eq:7}
\end{eqnarray}%
For simplicity, \pei{the single slit diffraction function has been ignored}. As shown in Eq.~(\ref{eq:7}), the second-order correlation function is a two-dimensional function of the detecting plane coordinates. In Fig.~\ref{entanglement}(1), the correlation function Eq.~(\ref{eq:7}) is plotted. This pattern is similar to the experimental result which has been obtained by Peeters \textit{et al} \cite{Peeters2009}.

As done in the analysis for thermal light, line (a) in Fig.~\ref{entanglement}(1) represents an interference pattern which has the same spatial resolution in the normal Young's double-slit experiment. The sub-wavelength result obtained by D'Angelo \cite{D'Angelo2001} can be rebuilt along line (b) in Fig.~\ref{entanglement}(1). The cross-section curves along these two lines are plotted in Fig.~\ref{entanglement}(2)(a)-(b). Lines (c) and (d) represent the scanning ways of the detectors with $(x_{1}=x, x_{2}=2x)$ and $(x_{1}=x, x_{2}=-x/2)$, respectively. In Fig.~\ref{entanglement}(2)(c)-(d), the cross-sectional curves corresponding to lines (c) and (d) are plotted, which peak spacings are $\lambda/3$ and $3\lambda$, respectively. As \pei{pointed out in} Ref.~\cite{Kothe2011}, \pei{the simultaneous two photons diffracted by a double-slit will propagate independently, so the photons in coincidence measurement are not limited in only one line but fully fill the whole detecting plane.}
Therefore, the one-dimensional cross-sectional curve deduced from the two-dimensional second-order correlation function may also have arbitrary narrow peak distances when the \pei{source is entangled light}.
\\
\\
\noindent \textbf{Discussion}

\noindent From the above discussion on both thermal and entangled light, \pei{we can see that a cross-sectional curve with }arbitrary spatial resolution can be built by properly choosing the scanning way of the detectors on the observing plane. However, cross sectional curves represent only partial information of the second-order interference. To get the full information, we should consider all \pei{possible combinations of arguments of the second-order correlation function. For example, including all possible variations of the two arguments,} the full picture of second-order interference of double-slit is a two-dimensional pattern.

For quantum lithography,
it is still challenging to realize a true laboratory demonstration due to the absence of the lithographic ($N$-photon absorption) materials, although proof-of-principle experiments that display certain aspects of quantum lithography have been performed. 
\pei{In these proof-of-principle experiments, $N$ photons should appear at same point on the $N$ detectors simultaneously because the $N$-photon absorber is simulated by $N$-fold coincidence.}
%In these proof-of-principle experiments, $N$-fold coincidence is utilized to simulate $N$-photon absorber, and $N$ photons should appear at same point on the $N$ detectors simultaneously if we compare the process of $N$-photon absorption on lithographic materials. 
From this point of view, we can see that if $N$ detectors scan in different paces and directions, it is impossible to take the advantage of the resulting super sub-wavelength patterns to achieve quantum lithography. In other words, for thermal light, the second-order interference pattern becomes flat when two detectors scanning in same directions ($x_1=x_2$), which clearly can not be used for writing sub-Rayleigh structures. While for entangled light, sub-wavelength interference pattern is gotten when scanning two detectors with $x_1=x_2$ (as it shown in Fig.~\ref{entanglement}(2)(b)). So entangled light source has the potential application in quantum lithography though the efficiency may be an obstacle \cite{Tsang2007,Kothe2011,Tsang08}. We also notice that the two-dimensional pattern measurement is not limited in space-momentum system. Such as, it can be used to the entropic entanglement measures in photon's orbital angular momentum \cite{Leach10}.

In summary, \pei{we recall the well-known fact that the second-order interference is in general a high-dimensional pattern rather than a one-dimensional curve.}
In the double-slit second-order interference experiments with either thermal or entangled light, when the two point detectors move along a certain way on the signal and reference detecting planes, the result obtained from the coincidence count of the detectors is only a cross-sectional curve of the two-dimensional interference pattern. In principle, one can build the cross-section curve of arbitrary spatial resolution only by choosing a proper scanning way of the two point detectors on the detecting planes. Furthermore, we find that this kind of super sub-wavelength interference of thermal light is impossible to be applied in quantum lithography. Our theory and experimental method for second-order interference may provide a deeper understanding of high-order correlations and \pei{hold promise in developing} their applications.
\\
\\
\noindent \textbf{Method}

\noindent There are two variables $x_1$ and $x_2$ in Eq.~(\ref{eq:6}), which represent the positions of two detectors, respectively. \pei{So full information of the second-order correlation function can be contained only by considering all different combinations of $x_1$ and $x_2$ in the detecting planes. In} the previous sub-wavelength interference experiments with a thermal light source \cite{Scarcelli2004,Xiong2005}, the $\lambda/2$ spatial resolution cross-sectional curves were obtained when two point detectors scanning in opposite directions ($x_{1}=x$, $x_{2}=-x$). However, there are no any mathematical and physical reasons to limit the scanning ways of two detectors with the coordinates $x_1$ and $x_2$ in Eq.~(\ref{eq:6}).
For example, if two point detectors move on the detecting planes in the way $(x_{1}=x$, $x_{2}=x/2)$, one can obtain the $2\lambda$ spatial resolution cross-sectional curve. In contrast to this case, if let the two point detectors move on the detecting planes in the way $(x_{1}=x$, $x_{2}=-2x)$, one can obtain the $\lambda/3$ spatial resolution cross-sectional curve. In general, the cross-sectional curve with the spatial resolution $N\lambda$ or $\frac{1}{N}\lambda$ (where $N > 0$) can be achieved when the two point detectors scan on the detecting planes in the way $(x_{1}=x$, $x_{2}=\frac{N-1}{N}x)$ or $(x_{1}=x$, $x_{2}=(1-N)x)$, respectively. It means that arbitrary spatial resolution cross-sectional curve can be built just by choosing an appropriate scanning way. This result seems surprising \pei{that, as a physical quantity, $G^{(2)}$ is not independent of the scanning ways}. However, noting the second-order correlation function Eq.~(\ref{eq:6}), we can find the reason why this arises\pei{---the }second-order correlation pattern should be a two-dimensional surface rather than a one-dimensional curve. In order to further clarify our argument, we perform a ghost interference experiment with similar setup to one employed in Ref.~\cite{Scarcelli2004,Xiong2005} using pseudo-thermal light \pei{as we described in the Results section}.

\noindent
\\
\noindent \textbf{Acknowledgements}

\noindent This work is supported by the Fundamental Research Funds for the
Central Universities, the National
Basic Research Program of China (Grant No. 2010CB923102) and the
National Natural Science Foundation of China (Grant Nos. 11004158, 11374008,
11074198, 11174233, 11074199 and 11374239).
\\
\\
\noindent \textbf{Author contributions}

\noindent R.L. constructed and operated the experiment, and collected the data. P.Z. and F.L. devised and designed the experiment. Y.Z. and H.G. carried out theoretical calculations. All authors contributed to the manuscript.
\\
\\
%\noindent  \textbf{Additional information}

%\noindent  The authors declare no competing financial interests. Reprints and permissions information is available online at http://npg.nature.com/reprintsandpermissions. Correspondence and requests for materials should be addressed to P.Z. or F.L. %\blk


\begin{thebibliography}{99}
\bibitem{Boto2000} Boto, A. N., Kok, P., Abrams, D. S., Braunstein, S. L., Williams, C. P., \& Dowling, J. P. Quantum interferometric optical lithography: exploiting entanglement to beat the diffraction limit. \textit{Phys. Rev. Lett.} \textbf{85}, 2733-2736 (2000).

\bibitem{Rayleigh} Rayleigh, L. Investigations in optics, with special reference to the spectroscope. \textit{Philos. Mag.} \textbf{8}, 261 (1879).

\bibitem{Boyd2012} Boyd, R. W., \& Dowling, J. P. Quantum lithography: status of the field. \textit{Quantum Inf. Process} \textbf{11}, 891-901 (2012).

\bibitem{explain} Actually, the Rayleigh limitation is smaller than $\lambda$. Here we use $\lambda$ for easy comparison to sub-wavelength.

\bibitem{D'Angelo2001} D'Angelo, M., Chekhova, M. V., \& Shih, Y. Two-photon diffraction and quantum lithography. \textit{Phys. Rev. Lett.} \textbf{87}, 013602 (2001).

\bibitem{Scarcelli2004} Scarcelli, G., Valencia, A., \& Shih, Y. Two-photon interference with thermal light. \textit{Europhys. Lett.} \textbf{68}, 618-624 (2004).

\bibitem{Xiong2005} Xiong, J., Cao, D. Z., Huang, F., Li, H. G., Sun, X. J., \& Wang, K. Experimental observation of classical sub-wavelength interference with thermal-like Light. \textit{Phys. Rev. Lett.} \textbf{94}, 173601 (2005).

\bibitem{Zhai2005} Zhai, Y. H., Chen, X. H., Zhang, D., \& Wu, L. A. Two-photon interference with true thermal light. \textit{Phys. Rev. A} \textbf{72}, 043805 (2005).

\bibitem{HBT19561} Brown, R. H., \& Twiss, R. Q. Correlation between photons in two coherent beams of light. \textit{Nature} \textbf{177}, 27-29 (1956).

\bibitem{HBT19562} Brown, R. H., \& Twiss, R. Q. A test of a new type of stellar interferometer on Sirius. \textit{Nature} \textbf{178}, 1046-1048 (1956).

\bibitem{HBT1974} Brown, R. H. Intensity Interferometer (Taylor Francis, London, 1974).

\bibitem{mandel} Mandel, L.,  \& Wolf, E. Photon Correlations Optical Coherence and Quantum Optics (Cambridge University Press, Cambridge, 1995).

\bibitem{Glauber1963} Glauber, R. J. Photon Correlations. \textit{Phys. Rev. Lett.} \textbf{10}, 84-86 (1963).

\bibitem{Glauber1963pr} Glauber, R. J. Coherent and Incoherent States of the Radiation Field. \textit{Phys. Rev.} \textbf{131}, 2766-2788 (1963).

\bibitem{Scarcelli2006} Scarcelli, G., Berardi, V., \& Shih, Y. Can two-photon correlation of chaotic light be considered as correlation of intensity fluctuations? \textit{Phys. Rev. Lett.} \textbf{96}, 063602 (2006).

\bibitem{Pittman1995} Pittman, T. B., Shih, Y. H., Strekalov, D. V., \& Sergienko, A. V. Optical imaging by means of two-photon quantum entanglement. \textit{Phys. Rev. A} \textbf{52}, R3429-R3432 (1995).

\bibitem{Cheng2004} Cheng, J., \& Han, S. Incoherent coincidence imaging and its applicability in X-ray diffraction. \textit{Phys. Rev. Lett.} \textbf{92}, 093903 (2004).

\bibitem{Gatti2004} Gatti, A., Brambilla, E., Bache, M., \& Lugiato, L. A. Ghost imaging with thermal light: comparing entanglement and classical correlation. \textit{Phys. Rev. Lett.} \textbf{93}, 093602 (2004).

\bibitem{QI2012} Boyd, R. W., \& Dowling, J. P. Special Issue: Quantum Imaging. \textit{Quantum Inf. Process} \textbf{11}, 887-1011 (2012).

\bibitem{phs94} Ribeiro, P. S., P\'adua, S., da Silva, J. M., \& Barbosa, G. A. Controlling the degree of visibility of Young's fringes with photon coincidence measurements. \textit{Phys. Rev. A} \textbf{49}, 4176-4179 (1994).

\bibitem{ejs99} Fonseca, E. J. S., Ribeiro, P. S., P\'adua, S., \& Monken, C. H. Quantum interference by a nonlocal double slit. \textit{Phys. Rev. A} \textbf{60}, 1530-1533 (1999).

\bibitem{EJS99} Fonseca, E. J. S., Monken, C. H., \& P\'adua, S. Measurement of the de Broglie wavelength of a multiphoton wave packet. \textit{Phys. Rev. Lett.} \textbf{82}, 2868-2871 (1999).

\bibitem{Sciarrino08} Sciarrino, F., Vitelli, C., De Martini, F., Glasser, R., Cable, H., \& Dowling, J. P. Experimental sub-Rayleigh resolution by an unseeded high-gain optical parametric amplifier for quantum lithography. \textit{Phys. Rev. A} \textbf{77}, 012324 (2008).

\bibitem{Tsang2009} Tsang, M. Quantum Imaging beyond the Diffraction Limit by Optical Centroid Measurements. \textit{Phys. Rev. Lett.} \textbf{102}, 253601 (2009).

\bibitem{Guerrieri2010} Guerrieri, F., Maccone, L., Wong, F. N., Shapiro, J. H., Tisa, S., \& Zappa, F. Sub-Rayleigh imaging via N-photon detection. \textit{Phys. Rev. Lett.} \textbf{105}, 163602 (2010).

\bibitem{Scarcelli} Scarcelli, G., Valencia, A., \& Shih, Y. Experimental study of the momentum correlation of a pseudothermal field in the photon-counting regime. \textit{Phys. Rev. A} \textbf{70}, 051802 (2004).

\bibitem{Steuernagel2004} Steuernagel, O. On the concentration behaviour of entangled photons. \textit{J. Opt. B: Quantum Semiclass. Opt.} \textbf{6}, S606-S609 (2004).

\bibitem{Tsang2007} Tsang, M. Relationship between resolution enhancement and multiphoton absorption rate in quantum lithography. \textit{Phys. Rev. A} \textbf{75}, 043813 (2007).

\bibitem{Kothe2011} Kothe, C., Bj\"ork, G., Inoue, S., \& Bourennane, M. On the efficiency of quantum lithography. \textit{New J. Phys.} \textbf{13}, 043028 (2011).

\bibitem{Peeters2009} Peeters, W. H., Renema, J. J., \& van Exter, M. P. Engineering of two-photon spatial quantum correlations behind a double slit. \textit{Phys. Rev. A} \textbf{79}, 043817 (2009).

\bibitem{Tsang08} Tsang, M. Fundamental Quantum Limit to the Multiphoton Absorption Rate for Monochromatic Light. \textit{Phys. Rev. Lett.}  \textbf{101}, 033602 (2008).

\bibitem{Leach10} Leach, J., Jack, B., Romero, J., Jha, A. K., Yao, A. M., Franke-Arnold, S., Ireland, D. G., Boyd, R. W., Barnett, S. M., \& Padgett, M. J. Quantum correlations in optical angle-orbital angular momentum variables. \textit{Science} \textbf{329}, 662 (2010).
\end{thebibliography}
\end{document}